\documentstyle[prl,aps,multicol,epsf]{revtex}
\begin{document}
\draft
\title{Does $m^*g^*$ diverge at a finite electron density in silicon inversion
layers?}
\author{M.P. Sarachik and S. A. Vitkalov}
\address{Physics Department, City College of the City
University of New York, New York, New York 10031}

\date{\today}
\maketitle

\begin{abstract}
For the two-dimensional electron system in silicon MOSFET's, the
scaled magnetoconductivity has been shown to exhibit critical behavior at
finite density $n_0$.  Analysis of these magnetotransport experiments
yields a product $g^*m^*$ that diverges at this density (here $g^*$
is the interaction-enhanced Land\'e $g$-factor and $m^*$ is the effective
mass).  This claim has been disputed based on direct determinations of 
the $g^*m^*$ obtained from Shubnikov-de Haas measurements.  We briefly
review these experiments, and possible sources of the discrepancies.

\end{abstract}


\begin{multicols}{2}

There is considerable current interest in the unusual behavior
of dilute two-dimensional systems of electrons (or holes)
\cite{revmodphys}.  Contrary to the long-held expectation that
non-interacting \cite{gang} or weakly interacting \cite{aal} electrons
become localized ({\it i. e.} are insulators) in two dimensions in the
limit of zero temperature, experiments within the past decade have shown
that the conductivity exhibits metallic temperature dependence within a
range of $low$ electron densities (where electron interactions are
actually quite strong).  Unexpected metallic behavior has been observed
in this low-density regime: for electron ($n_s$) or hole ($p_s$)
densities above some critical density $n_c$ (or $p_c$), the conductivity
$\sigma$ increases with decreasing temperature; $\sigma$ is approximately
independent of temperature near the quantum unit of conductance
($e^2/h \approx 3.9 \times 10^{-5} \Omega^{-1}$) at $n_c$ and exhibits
insulating behavior below this critical density \cite{krav}.  In silicon
MOSFETs, the conductivity has been shown to increase down to a temperature
$\approx 35$ mK for electron densities just above $n_c$ \cite{kravlowT}. 
This suggests there exists a metallic phase and a true metal-insulator
transition in dilute strongly interacting electron (hole) systems.

The response of these systems to external magnetic field is unusual and 
dramatic.  An external magnetic field of the order of a few Tesla applied
parallel to the plane of the electrons gives rise to an enormous positive
magnetoresistance \cite{dolgopolov,simonian,pudalov,yoon} on both sides
of the transition: the longitudinal resistivity increases (conductivity
decreases) dramatically as a function of magnetic field $H$ applied
parallel to the plane of the electrons and saturates to a value that is
approximately constant for magnetic fields $H > H_{sat}$ with the value
of $H_{sat}$ depending on the electron (hole) density $(n_s)$. 
Interestingly, a parallel magnetic field has been shown to suppress the
metallic behavior \cite{simonian,Kevin}.  Moreover, recent experiments
have yielded evidence that the magnetotransport exhibits critical
behavior at a finite density \cite{ferro,Shashkin}.  It is not clear
whether and how the metallic temperature-dependence and the magnetic
response are related.

These developments have elicited intense interest, and have fueled a 
lively debate.  The issue is whether the novel effects found in dilute
two-dimensional materials represent fundamentally new physics or whether
they can be explained by an extension of physics that is already
understood.  A view held by some is that these features signal a true
zero-temperature quantum phase transition to a novel ground state at T=0
(such as a ``perfect'' metal, a superconductor, a ferromagnet, a spin 
liquid, a Wigner glass, etc. \cite{revmodphys}).  Others believe that
the metallic temperature dependence can be understood within the
framework of Fermi liquid theory and is due, for example, to
temperature-dependent screening, percolation, interband scattering, or
scattering at charged traps.  No consensus has been reached; the
enigmatic behavior of dilute two-dimensional systems continues to
be one of the most interesting unresolved issues in Condensed Matter
Physics.

\section{Does $g^*m^*$ diverge in silicon MOSFET's?}

For the 2D electron system in silicon MOSFET's, several groups have
determined renormalized values of the effective mass $m^*$ and the
Land\'e $g$-factor $g^*$ based on measurements of the magnetoconductivity
and of Shubnikov-de Haas oscillations as a function of temperature and
electron density.  There is currently disagreement whether either or both
of these renormalized quantities diverge at a finite electron density in
this system \cite{shayegan}.  This is indeed an important question: such a
divergence would signal critical behavior of the spin susceptibility in
silicon MOSFET's. In this paper we briefly review and examine this
controversy.

A quantitative characterization of the system's response to parallel
magnetic field has been difficult to obtain.  Attempts to scale the
experimentally measured magnetoresistance yield a match at low fields or
at high magnetic fields, but none provide satisfactory results over the
entire range, particularly in the low-density regime.  By considering the
magnetoconductance instead of the magnetoresistance, we have recently
succeeded in obtaining an excellent data collapse over a 
\vbox{
\vspace{0in}
\hbox{
\hspace{-0.2in}
\epsfxsize 5 in \epsfbox{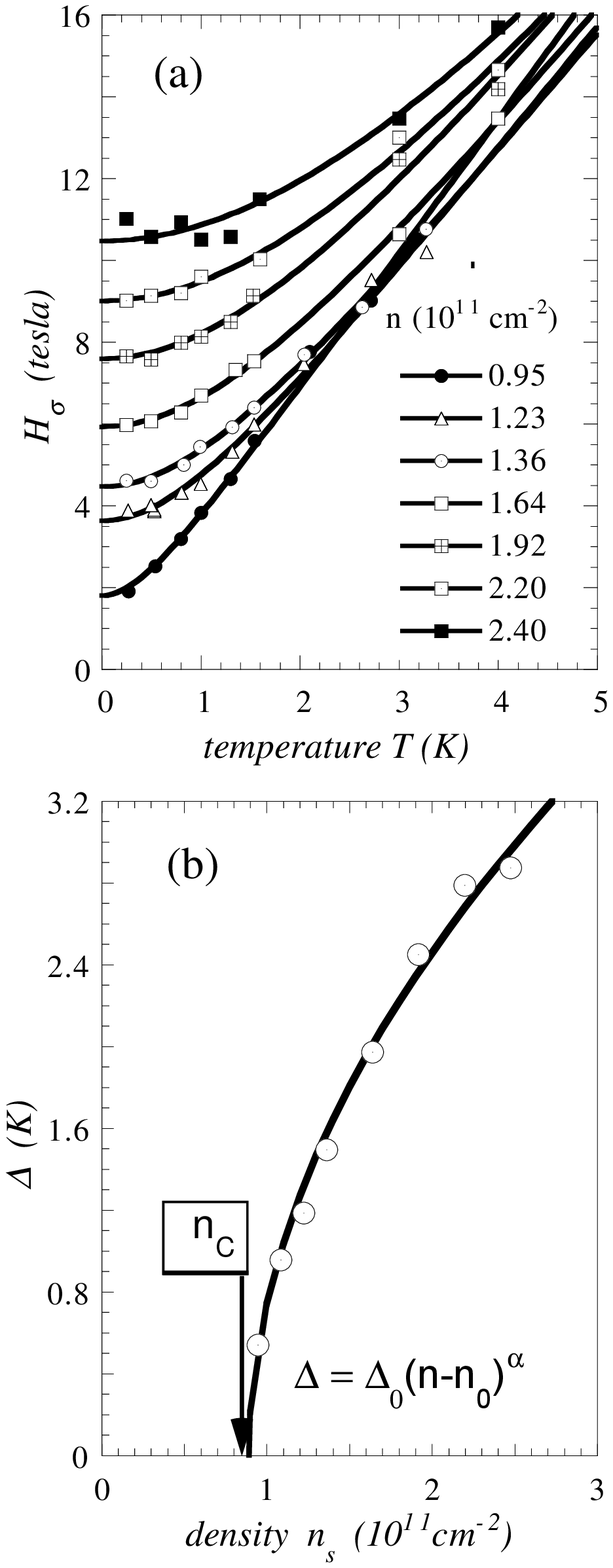}
}
}
\vskip 0cm
\refstepcounter{figure}
\parbox[b]{3.1in}{\baselineskip=12pt FIG.~\thefigure.
(a) $H_{\sigma}$ versus temperature for different electron
densities; the solid lines are fits to the empirical form
$H_\sigma(n_s,T)=A (n_s) [[\Delta(n_s)]^2 +T^2]^{1/2}$.
(b) The parameter $\Delta$ as a function of electron density; the solid
line is a fit to the expression $\Delta = \Delta_0 (n_s - n_0)^\alpha$.
\vspace{0.10in}
}
\label{fig1}
broad range of
electron densities and temperatures using a single scaling parameter
$H_\sigma$ \cite{ferro}.  (Possible reasons for the difference between
resistivity and conductivity are discussed below).  We showed that the
scaling parameter $H_\sigma$ is consistent with the empirical relation:
$H_\sigma(n_s,T)=A (n_s) [[\Delta(n_s)]^2 +T^2]^{1/2}$.  Fig. 1 (a) shows
$H_\sigma$ versus $T$ for different densities $n_s$ approaching the
transition.   Fits to this expression yield values of $\Delta$ shown in
Fig. 1 (b) as a function of electron density $n_s$.

\vbox{
\vspace{0in}
\hbox{
\hspace{-0.2in}
\epsfxsize 5 in \epsfbox{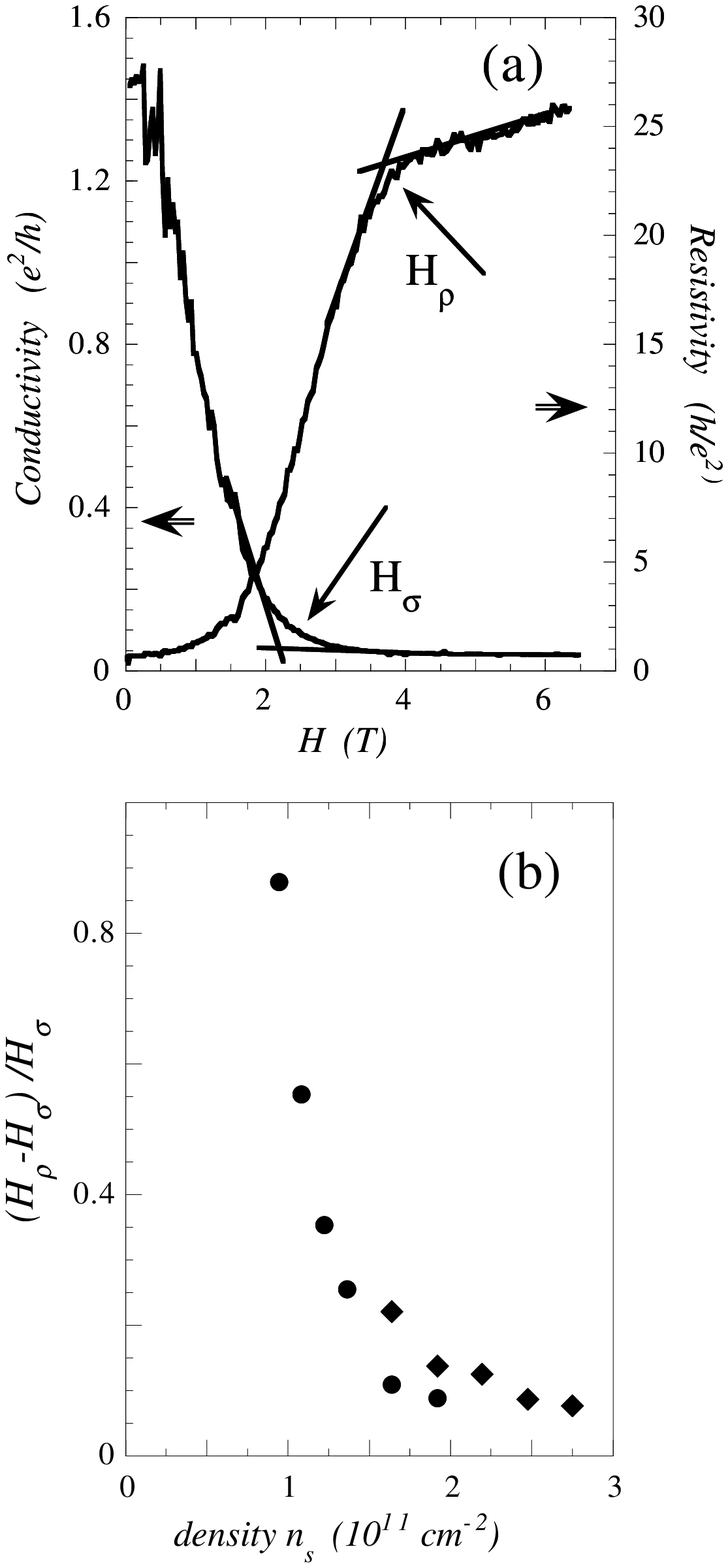}
}
}
\vskip 0.5cm
\refstepcounter{figure}
\parbox[b]{3.1in}{\baselineskip=12pt FIG.~\thefigure.
(a) The conductivity (left axis, and resistivity (right axis) as
a function of magnetic field applied parallel to the electron plane; the
electron density $n_s = 0.94 \times 10^{11}$ cm$^{-2}$; $H_\sigma$ and
$H_\rho$ denote the saturation fields deduced from plotting the
conductivity and resistivity, respectively.  (b) The (normalized)
difference between $H_\sigma$ and $H_\rho$ versus electron density. 
(Different symbols denote two different samples.
\vspace{0.10in}
}
\label{fig2}

The parameter $\Delta$ represents an energy $k_B \Delta$: for high
densities and low temperature, $T<\Delta \sim \hbar/\tau_H$, $H_{\sigma}$
is determined by $\Delta$ and the system is in the zero temperature
limit; at lower densities the measurement temperature $T>\Delta \sim
\hbar/\tau_H$, the field $H_{\sigma}$ is dominated by thermal effects. 
The energy scale $k_B \Delta$ depends on electron density and
extrapolates to zero at a finite density $n_0 \approx 0.85 \times
10^{11}$  cm$^{-2}$.  That we have identified a critical regime and an
approach to $\Delta = 0$ is further supported by our finding that for a
density $n_s \approx n_0$ the magnetoconductivity scales with $H/T$ down
to our lowest measuring temperature ($0.25$ K), indicating that
$\Delta=0$ in the vicinity of $n_0$.  The observed  critical behavior of
the magnetoconductivity suggests there is a (zero-temperature)
quantum phase transition at a density $n_0$ \cite{ferro}.

The origin of the critical behavior of $H_\sigma$ is currently under
investigation.  Based on early observations at high electron densities
\cite{okamoto,vitkalov,angular} that the parameter
$H_\sigma$ is close to the field required for complete spin polarization
of the 2D carriers, we suggested that the critical behavior of the
magnetoconductivity is associated with the spin degrees of freedom of
the system \cite{ferro}.  In the remainder of this paper, we compare
values of $g^*m^*$ deduced from the magnetoconductivity with those
obtained from Shubnikov-de Haas measurements.

\vbox{
\vspace{0.4 in}
\hbox{
\hspace{-0.2in}
\epsfxsize 3.4 in \epsfbox{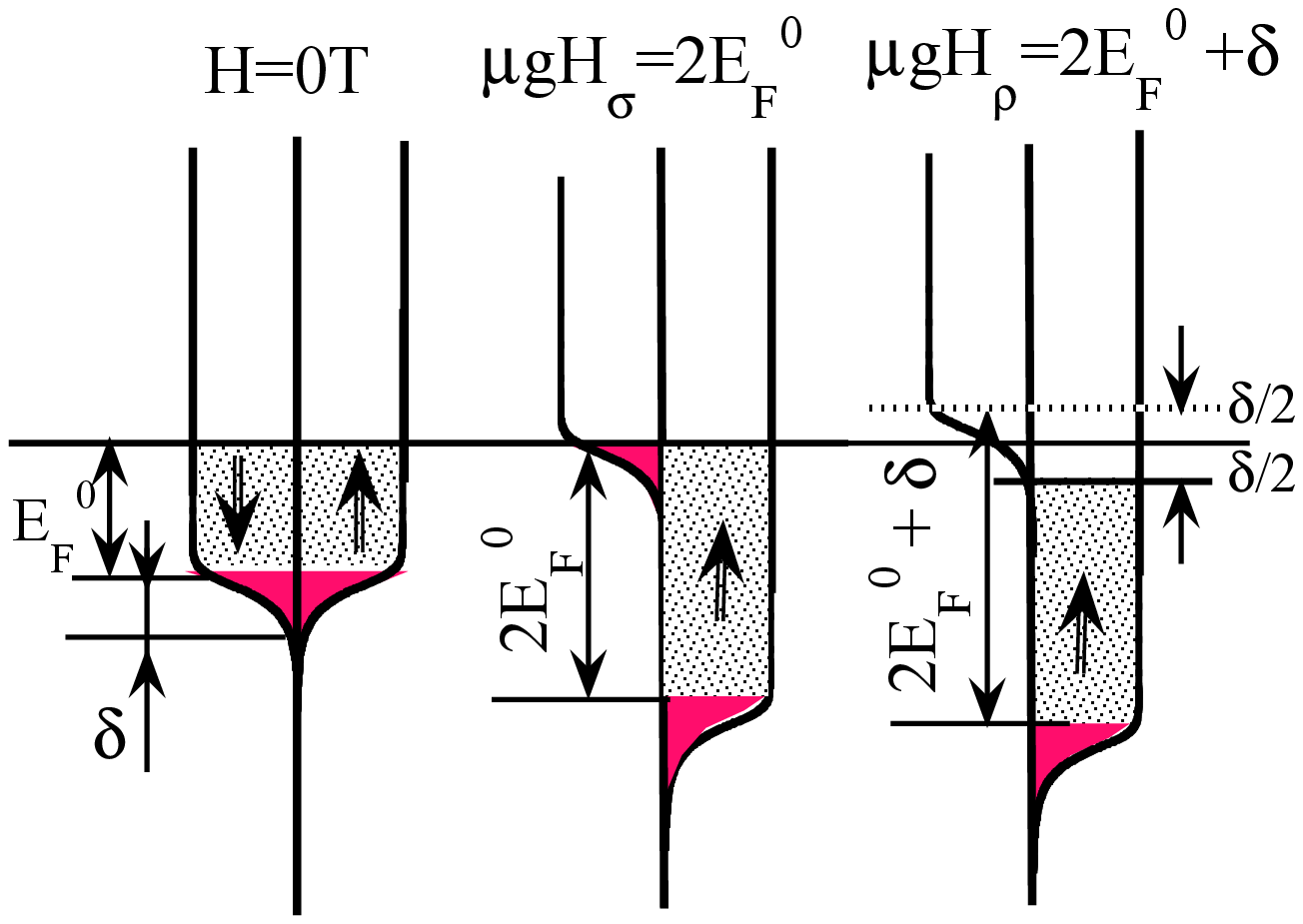}
}
}
\vskip 0.5cm
\refstepcounter{figure}
\parbox[b]{3.1in}{\baselineskip=12pt FIG.~\thefigure.
Schematic diagram of the bands in silicon MOSFET's illustrating
the role of band tails: in zero field, for $H=H_{\sigma}$, and at
$H=H_{\rho}$.  See text for discussion.
\vspace{0.10in}
}
\label{fig3}

For high electron densities, Shubnikov-de~Haas measurements
\cite{okamoto,vitkalov,angular,tutuc} have shown directly that the
saturation of the magnetoresistance at field $H_{sat}$ is associated with
full polarization of the electron spins by the in-plane magnetic field. 
It is now generally recognized that for lower densities, the relation
between full polarization and resistivity saturation is considerably more
complicated.  As shown in Fig. 2 (a), we have recently noted that
examination of the resistivity and of the conductivity (its inverse), do
not yield the same apparent saturation point.  As illustrated in Fig. 3,
we have suggested \cite{gm} that the field $H_\sigma$ where the
conductivity appears to saturate corresponds to full polarization of the
highly mobile band states $(g^* \mu_B H_\sigma = 2 E_F$), while the
higher magnetic field $H_\rho$ required for resistivity saturation
corresponds to full alignment of ALL the electrons, including those in
tail states below the bottom of the band ($(g^* \mu_B H_\rho = 2 E_F +
\delta$).  As shown in Fig. 2 (b), the difference between $H_\sigma$ and
$H_\rho$ becomes larger as the density is decreased toward $n_0$,
indicating that disorder plays an increasingly important role.

We suggest that the $H_{sat}$ deduced from our magnetoconductance
measurements is associated with the response of the highly mobile band
states (and not the localized states in the band tails) so that $H_\sigma
= H_{sat}$.  If one sets $g^*\mu_BH_{sat}=2E_F$, one can obtain
zero-temperature values of the susceptibility $\chi^* \propto g^*m^*$. 
The open circles plotted in Fig. 4 denote the inverse of the
normalized susceptibility $\chi_0/\chi^*$ plotted as a function of
electron density $n_s$.

\vbox{
\vspace{0.4 in}
\hbox{
\hspace{-0.2in}
\epsfxsize 3.2 in \epsfbox{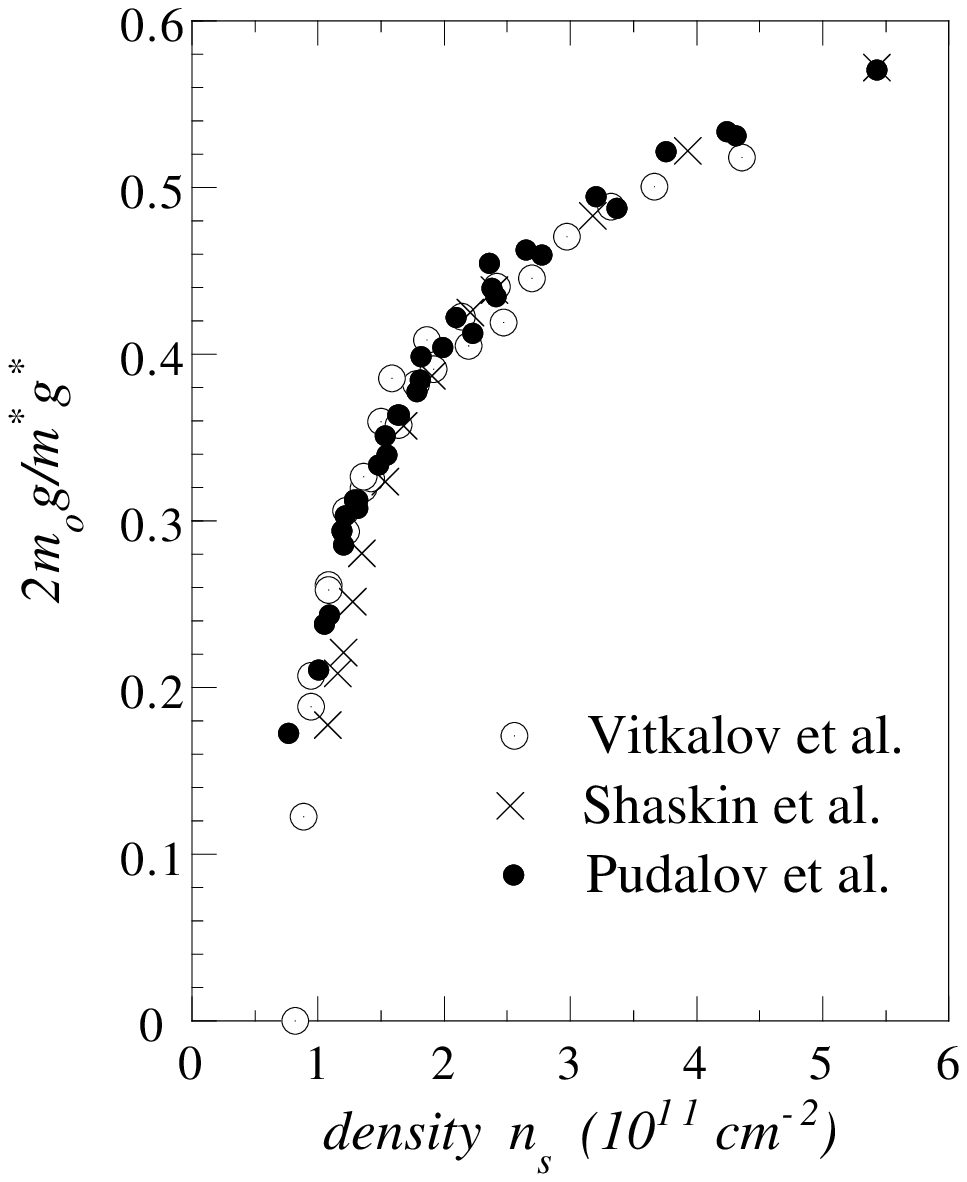}
}
}
\vskip 0.5cm
\refstepcounter{figure}
\parbox[b]{3.1in}{\baselineskip=12pt FIG.~\thefigure.
The normalized inverse susceptibility $\chi_0/\chi^*$ versus
electron density deduced by three different methods (see discussion in
the text).
\vspace{0.10in}
}
\label{fig4}

As mentioned earlier, it is not possible to scale the magnetoresistance
data over the full range of magnetic field.  Shashkin {\it et al.}
\cite{Shashkin} scaled the magnetoresistance with $H/H_c$ by matching the
low-field behavior of the magnetoresistivity; this entails a considerable
mismatch in large fields.  Interestingly, the low field response
is a measure of the high-mobility band states so that the $H_c$ obtained
by Shashkin {\it et al.} is similar to our $H_\sigma$ by the arguments
given above.  Shashkin {\it et al.}'s findings differ from ours in some
respects: for example, they find that $H_c$ extrapolates $linearly$ to
zero at a finite density; for the sample used in their experiment they
found $n_0 = n_c$.  Despite these differences, however, their results for
$\chi_0/\chi^*$, designated by the crosses in Fig. 4, are quite similar
to ours.  The values deduced by both groups from transport measurements
indicate that $(g^*m^*)^{-1}$ extrapolates to zero ($g^*m^*$ diverges) in
silicon MOSFET's at a finite density.

Pudalov {\it et al.} \cite{Pudalov} dispute this claim based on direct
determinations of $g^*m^*$ from Shubnikov-de Haas (SdH) measurements in
crossed magnetic fields down to very low densities.  The susceptibility
$\chi_0/\chi^*$ determined from these measurements is shown by the closed
circles in Fig. 4.  The susceptibility obtained by all three groups, two
using different analyses of magnetotransport and the third from SdH
measurements, are remarkably similar.  However, Pudalov {\it et al.}
report that detailed examination of their data reveals a sharp increase
for the value of $g^*m^*$ (by a factor of $\approx 4.6$) as the electron
density decreases toward $n_c \approx 0.85 \times 10^{11}$ cm$^{-2}$, but
no divergence at this density.  These conflicting claims must be resolved
through further experimentation, and a better understanding of both
transport and Shubnikov-de Haas data.

\vbox{
\vspace{0in}
\hbox{
\hspace{-0.2in}
\epsfxsize 4.8 in \epsfbox{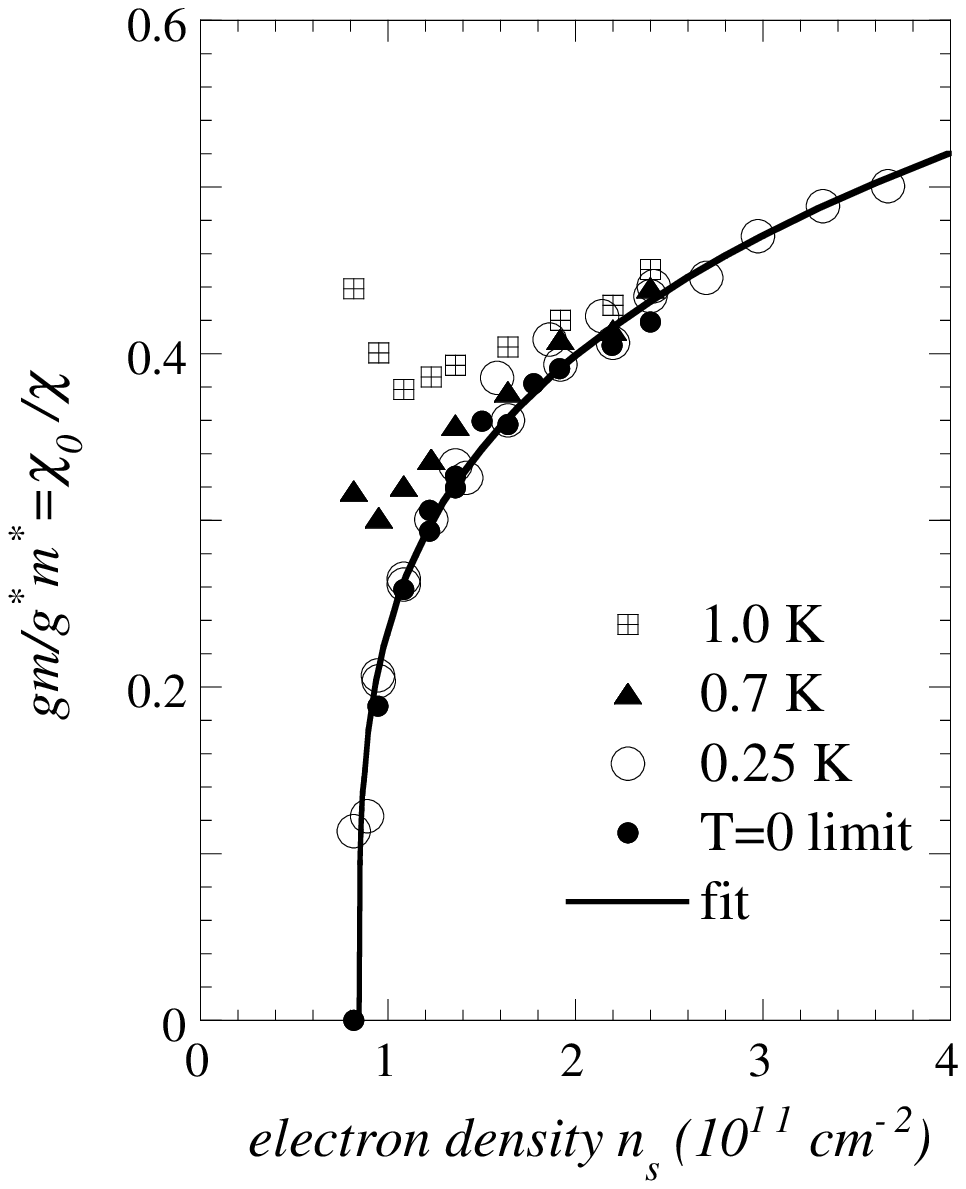}
}
}
\vskip -0.3cm
\refstepcounter{figure}
\parbox[b]{3.1in}{\baselineskip=12pt FIG.~\thefigure.
The normalized inverse susceptibility $\chi_0/\chi^*$ versus
electron density at $0.25$ K, $0.7$ K, and $1$ K.  The closed circles
denote extrapolations to $T=0$.
\vspace{0.10in}
}
\label{fig5}

Our procedure for scaling the magnetoconductance yields a parameter
$H_\sigma$ from which we have deduced the renormalized susceptibility
$(g^*m^* \propto \chi^*)$, and the zero-temperature limit of this
quantity appears to diverge.  The $H_\sigma$ that results from this
scaling ansatz may not provide a reliable measure of the susceptibility,
particularly at very low densities where disorder and band-tail states
play an important role.  By the same token, it is important to take
additional, very precise measurements of the Shubnikov-de Haas
oscillations.  Fig. 1(a) shows that $H_\sigma$ depends on temperature. 
However, if one restricts one's attention to the region below $1$ K,
$H_\sigma$ depends on temperature $only$ for low densities near the
apparent divergence at $n_0$.  The importance of examining the
temperature dependence of $\chi^*$ and obtaining a reliable extrapolation
to $T=0$ is further illustrated in Fig. 5, where the normalized
susceptibility $\chi_0/\chi^*$ obtained from our data is plotted
at different temperatures between $0.25$ K and $1$ K.  If it occurs,
the divergence of $\chi$ is apparent only in the limit of zero
temperature.  This possibility is further supported by recent experiments
of Reznikov and Sivan \cite{reznikov}, who have determined the
magnetization by measuring the change in chemical potential with applied
magnetic field applying the thermodynamic relation $d\mu/dH = - dM/dn_s$
(here $\mu$ is the chemical potential, $H$ is the magnetic field, $M$ is
the magnetization, and $n_s$ is the electron density).  Their preliminary
results show that the susceptibility depends on temperature, and the
$T$-dependence appears to extend to lower temperatures at low electron
densities.  This is consistent with the temperature dependence of $\chi^*$
deduced from our transport measurements shown in Fig. 5.  On the other
hand, Pudalov {\it et al.} \cite{Pudalov} report that the $(g^*m^* \propto
\chi^*)$ obtained from their Shubnikov-de Haas measurements is almost
independent of temperature over the range $0.3 - 1$ K.  Additional,
detailed studies of Shubnikov-de Haas are needed at low electron
densities near $n_0$ to determine whether $g^*m^*$ is indeed independent
of temperature at very low temperatures, or whether the behavior is
closer to that shown in Fig. 1 (a) and Fig. 5.  We note that low density
is precisely the region where SdH measurements are most difficult to
perform and interpret.

\section*{Acknowledgements}

We thank V. M. Pudalov, M. Gershenson and S. V. Kravchenko for providing
data for Fig. 4.  This work was supported by the US Department of
Energy under Grant No.~DE-FG02-84ER45153 and National Science Foundation
grant DMR-0129581.

\end{multicols}
\end{document}